\documentclass[preprint,showpacs,preprintnumbers,amsmath,amssymb]{revtex4}

\usepackage{graphicx}
\usepackage{epsfig}		
\usepackage{dcolumn}
\usepackage{bm}

\def\0{\mbox{\tiny $0$}}
\def\1{\mbox{\tiny $1$}}
\def\2{\mbox{\tiny $2$}}
\def\3{\mbox{\tiny $3$}}
\def\4{\mbox{\tiny $4$}}
\def\5{\mbox{\tiny $5$}}
\def\6{\mbox{\tiny $6$}}
\def\7{\mbox{\tiny $7$}}
\def\8{\mbox{\tiny $8$}}
\def\9{\mbox{\tiny $9$}}
\def\a{\mbox{\tiny $\alpha$}}
\def\b{\mbox{\tiny $\beta$}}

\def\n{\mbox{\tiny $n$}}
\def\k{\mbox{\tiny $k$}}

\def\f14{\mbox{\tiny $\frac{1}{4}$}}

\def\L{\mbox{\tiny $L$}}

\def\ii{\mbox{\tiny $i$}}
\def\l{\mbox{\tiny $l$}}

\def\a{\mbox{\tiny $a$}}
\def\P{\mbox{\tiny $P$}}

\def\j{\mbox{\tiny $j$}}
\def\mi{\mbox{\tiny $-$}}

\def\bb#1{\mbox{\footnotesize $(#1)$}}

\begin{document}

\title{Lorentz-violating dilatations in the momentum space and some extensions on non-linear actions of Lorentz algebra-preserving systems}


\author{A. E. Bernardini}
\email{alexeb@ifi.unicamp.br}
\affiliation{Instituto de F\'{\i}sica Gleb Wataghin, UNICAMP,\\
PO Box 6165, 13083-970, Campinas, SP, Brasil.}
\author{R. da Rocha}
\email{roldao@ifi.unicamp.br}
\affiliation{Instituto de F\'{\i}sica Gleb Wataghin, UNICAMP,\\
PO Box 6165, 13083-970, Campinas, SP, Brasil\\
Instituto de F\'{\i}sica Te\'{o}rica, UNESP,\\
Rua Pamplona 145, 01405-900, S\~{a}o Paulo, SP, Brasil.}

\date{\today}

\begin{abstract}
We work on some general extensions of the formalism for theories which preserve the relativity of inertial frames with a nonlinear action of the Lorentz transformations on momentum space.
Relativistic particle models invariant under the corresponding deformed symmetries are presented with particular emphasis on deformed dilatation transformations.
The algebraic transformations relating the deformed symmetries with the usual (undeformed) ones are provided in order to preserve the Lorentz algebra.
Two distinct cases are considered: a deformed dilatation transformation with a spacelike preferred direction and a very special relativity embedding with a lightlike preferred direction.
In both analysis we consider the possibility of introducing quantum deformations of the corresponding symmetries such that 
the spacetime coordinates can be reconstructed and the particular form of the real space-momentum commutator remains covariant.
Eventually feasible experiments, for which the non-linear Lorentz dilatation effects here pointed out may be detectable, are suggested.
\end{abstract}

\pacs{03.30.+p, 11.30.Cp, 11.30.-j}
\keywords{Dilatations - Lorentz Trasnformations - Very Special Relativity - Dirac Equation}
\date{\today}

\maketitle

\section{Introduction}

Theoretical issues pointing to the possibility that many empirical successes of
special relativity do need not demand Lorentz invariance of the underlying theory have been widely discussed in several frameworks \cite{Ame01,Mag02,Kos06,AmeDSR}.  
The characteristic scale of such theories is likely to be associated with the Planck's energy $E_{\P\l} \sim 10^{\1\9}$ GeV.
However, the current attainable energies are minuscule compared to this scale, so that experimental signals are expected to be heavily suppressed.

Among the most sensitive experimental data, from which such an assertion can eventually be confirmed, there are the threshold anomalies in the ultra high energy cosmic ray (UHECR) protons \cite{Ber99,Bie01,Tak98}, and $TeV$ photons \cite{Fin01} supposedly explained by modifications of the energy momentum dispersion relations \cite{Ame01,Mag02,Mag03}.
In this context, and also motivated by ideas from quantum gravity \cite{Smo95,Ame02}, an extension of special theory of relativity, known as deformed (or doubly) special relativity (DSR) \cite{AmeDSR}, was firstly proposed in manner to introduce two independent scales: the Planck scale and the velocity of light.
These theoretical advances have been paralleled by a perturbative framework developed to investigate a certain class of departures from Lorentz invariance.
For instance, Coleman and Glashow \cite{Col99,Col97} suggested spacetime translations along with exact rotational symmetry in the rest frame of the cosmic background radiation, but allow small departures from boost invariance in this frame. 
In some other cases the introduction in the Lagrangian of more general Lorentz-violating terms are analyzed \cite{Kos97,Kos98,DeG06}. 
Although no decisive evidence contradicting the exact Lorentz invariance has yet been experimentally detected, ever more sensitive searches are being carried out.
In particular, Cohen and Glashow pursue a different approach to the possible failure of Lorentz symmetry denominated {\em very special relativity} (VSR) \cite{Gla06A,Gla06B}, which has been expanded for studying some peculiar aspects of neutrino physics.
The VSR is based on the hypothesis that the spacetime symmetry group of nature is smaller than the Poincar\'{e} group, consisting of spacetime translations and one of certain subgroups of the Lorentz group.
In certain sense, it introduces a preferred direction $n_{\mu}$.

In this manuscript we present two relativistic particle models invariant under the corresponding deformed symmetries with particular emphasis on deformed dilatation transformations.
We intend to distinctly analyze two particular choices of preferred directions embedded in the framework(s) with Lorentz invariance violation.
The meaning of {\em Lorentz-violating} here adopted can be ``attenuated'' in the sense that the theory
and its resultant transformations preserve the Lorentz algebra. 
It just refers to the introduction of a preferred frame and to the modification of the algebra Casimir invariant which results from
the nonlinear action of the Lorentz group on momentum space.
In the first part of the manuscript, we implement a generalized deformed dilatation transformation from which we
notice that the non-linear Lorentz transformations with an invariant energy scale can be obtained with particular choice of invariant spacelike preferred direction.
Suitably changing the focus, but keeping at the same framework, we obtain the dispersion relation and the equation of motion for a
propagating fermionic particle by examining some previous claims for a preferred lightlike direction embedded in the framework of very special relativity (VSR).
In both analysis we examine the possibility of introducing quantum deformations of the corresponding symmetries such that 
the spacetime coordinates can be reconstructed and the particular form of the commutator remains covariant.

Since we treat some specific proposals to define nonlinear actions of the Lorentz group while preserving the Lorentz algebra, it can be said that the study fits in the general framework of Doubly Special Relativity (DSR) \cite{Gho06,Kow02,Hin05}.
In section II we are concentrated on the analysis of (deformed) dilatation transformations embedded in LV systems.
In particular, we establish a deformed dilatation transformation by introducing a preferred direction $n_{\mu}$, which however preserves the Lorentz algebra. 
Once the deformed symmetries have been defined in the momentum space, the next issue concerns the covariant formulation of the real space coordinate transformation as well as the determination of the commutation relations.
Finally, we notice that the LV system transformations with an invariant energy scale obtained in \cite{Mag02} correspond to a particular choice of invariant spacelike preferred direction of a generalized deformed dilatation transformation.
The second step of our study presented in section III is concerned with the obtention of the equation of motion for a Dirac particle in
an extension of the general class of DSR theories \cite{Gho06,Kow02}.
By keeping the focus on the same framework where the Lorentz algebra is preserved, we examine some previous claims for a preferred axis at $n_{\mu}$($\equiv(1,0,0,1)$), $n^{2}=0$, embedded in the framework of VSR.
We show that, in a relatively high energy scale, the corresponding equation of motion is reduced to a conserving lepton number chiral equation previously predicted in the literature.
On the contrary, in a relatively low energy scale, the equation is reduced to the usual Dirac equation for a free propagating fermionic particle.
It is accomplished by the suggestive analysis of some special cases where a nonlinear modification of the action of the Lorentz group is generated by the addition of a modified conformal transformation which, meanwhile, preserves the structure of the ordinary Lorentz algebra in a very peculiar way. 
We draw our conclusions in section IV.

\section{Dilatation transformations}

It has been know since the first half of the last century that the equations of the electromagnetic theory, the most appropriate instrument for such analysis, are invariant under the 15-parameter group of transformations called the conformal group \cite{1910}, which 
contains the 10-parameter Poincar\'{e} group as a subgroup.
The remaining five parameters are associated with an one parameter group of dilatations,
\small\begin{equation}
p_{\mu} \rightarrow \lambda p_{\mu},
\end{equation}\normalsize
and the following 4-parameter group of transformations
\small\begin{equation}
p_{\mu} \rightarrow \frac{p_{\mu} + p^{\2}\eta_{\mu}}{1 + 2\,p_{\nu} \eta^{\nu} + \eta^{\2}p^{\2}}.
\end{equation}\normalsize
The reason that the Poincar\'{e} transformations have received much more attention than the other transformations of the conformal group is that, as soon as massive particles are introduced into the theory, the conformal invariance is reduced to the invariance with respect to the Poincar\'{e} transformations.
Strictly speaking, the above arguments is an exaggeration of the facts.
Conformal symmetries are equally affected for massless models and it has been demonstrated that dilatation invariance may be exhibited 
within the same conceptual framework as that within which one usually discusses Lorentz invariance \cite{Gar70}.

In general lines, there have been some crucial issues related to the application of conformal transformations, and more specifically dilatations, to physical models.
Traditionally, these generalizations have been rejected as being physically uninterpretable since this naturally requires that one leaves the confines of the special theory of relativity.
However it might be possible to introduce quantum deformations of these symmetries, such that the particular form of the space-momentum commutator remains covariant.
In fact, the covariant prescription to define the spacetime coordinates has already been proposed in the literature \cite{Mig03} 
In this sense, the derivation of these properties as well as the determination of the commutation relations are the main results of this section.
For this aim we establish a deformed dilatation transformation that transforms $p_{\mu}$ as
\small\begin{equation}
p_{\mu} \rightarrow \frac{p_{\mu}}{1 - \lambda \,p_{\nu} n^{\nu}},
\end{equation}\normalsize
by introducing a preferred direction $n_{\mu}$, which however preserves the Lorentz algebra. 
The resultant formulation, with the assumption of an invariant spacelike preferred direction embedded in the deformed dilatation transformation,
allows us to recover the results of the formulation of generalized Lorentz invariance with an invariant energy scale expressed in \cite{Mag02,Mag03}.

\subsection{Embedding an extension of dilatation transformations in LV systems}

Let us start with the definition of the momentum space $\mathcal{M}$ as the four-dimensional vector space consisting of momentum vectors $p_{\mu}$.
The ordinary Lorentz generators act as
\small\begin{equation}
L_{\mu\nu} = p_{\mu}\partial_{\nu} - p_{\nu}\partial_{\mu}
\label{pp02}
\end{equation}\normalsize
where $\partial_{\mu} \equiv \partial/\partial p^{\mu}$, and we assume the Minkowski metric signature and that all generators are anti-Hermitian (also $\mu,\,\nu = 0,\,1,\,2,\,3$, and $i,\,j,\,k = 1,\,2,\,3$ and the velocity of the light $c = 1$).
In adition, the dilatation generator $D \equiv p_{\nu}\partial^{\nu}$ acts on momentum space as $D\circ p_{\mu} = p_{\mu}$. 
The ordinary Lorentz algebra is constructed in terms of the usual rotations $J^{\ii}\equiv \epsilon^{\ii\j\k}L_{\j\k}$ and boosts $K^{\ii} \equiv L^{\ii\0}$ as
\small\begin{equation}
[J^{\ii}, K^{\j}] = \epsilon^{\ii\j\k}K_{\k};~~~~[J^{\ii}, J^{\j}] = [K^{\ii}, K^{\j}] = \epsilon^{\ii\j\k}J_{\k}
\label{pp02A}
\end{equation}\normalsize
where $\epsilon^{\ii\j\k}$ is the antisymmetric tensor by index permutation with $\epsilon^{\1\2\3} = 1$.
In order to introduce the nonlinear action that modifies the ordinary Lorentz generators but, however,
preserves the structure of the algebra, we suggest the following {\em ansatz} for a generalized transformation, 
\small\begin{equation}
\mathcal{D} \equiv \lambda \,p_{\nu}n^{\nu} \,p_{\a}\partial^{\a}
\label{pp03}
\end{equation}\normalsize
which acts on the momentum space as 
\small\begin{equation}
\mathcal{D} \circ p_{\mu} \equiv \lambda \, p_{\nu}n^{\nu} \,p_{\mu}
\label{pp04}
\end{equation}\normalsize
where $\lambda$ is a physical constant (eventually, the Planck's length) and $n_{\nu}$ is a constant vector describing the preferred direction.
These constants encode the physical properties of this transformation.

We assume the new action can be considered to be a nonstandard and nonlinear embedding of the Lorentz group into a modified conformal group which, as we shall notice in the following for the case of main interest, despite the modifications, satisfies precisely the ordinary Lorentz algebra (\ref{pp02A}).
To exponentiate the new action, we note that 
\small\begin{equation}
k^{\ii} = U^{^{\mi 1}}\hspace{-0.35 cm}\bb{\mathcal{D}}\, K^{\ii}\, U\bb{\mathcal{D}} ~~\mbox{and}
~~ j^{\ii} = U^{^{\mi 1}}\hspace{-0.35 cm}\bb{\mathcal{D}} \,J^{\ii} \,U\bb{\mathcal{D}}
\label{pp06}
\end{equation}\normalsize
where the $p_{\mu}$-dependent transformation $U\bb{\mathcal{D}}$ is given by
$U\bb{\mathcal{D}\bb{\lambda, p_{\mu}}} \equiv{\exp[\mathcal{D}\bb{\lambda, p_{\mu}}]}$.
The nonlinear representation is then generated by $U\bb{\mathcal{D}\bb{\lambda, p_{\mu}}} \circ p_{\mu}$ and,
despite not being unitary ($U\bb{\mathcal{D}\bb{\lambda, p_{\mu}}} \circ p_{\mu} \neq p_{\mu}$),
it has to preserve the structure of the algebra, which is guaranteed by the easily verifiable commuting
relation\footnote{In fact, the Eq. (9) is not a necessary condition to preserve the algebra of the $L^{\mu\nu}$ generators (Lorentz algebra).
In the problem that we are analyzing, however, it can be demonstrated that it is a sufficient condition for preserving the Lorentz algebra 
since, after some extensive mathematical manipulations, we can verify that $[M_{\mu\nu}, M_{\a\b}] = [L_{\mu\nu}, L_{\a\b}] + [[L_{\mu\nu}, L_{\a\b}], \mathcal{D}]$
if $[[L_{\mu\nu},\,\mathcal{D}\bb{p_{\mu}}],\,\mathcal{D}\bb{p_{\mu}}] = 0$ when $M_{\mu\nu} =  L_{\mu\nu} + [L_{\mu\nu}, \mathcal{D}]$.}, 
\small\begin{equation}
\left[[L_{\mu\nu},\,\mathcal{D}\bb{p_{\mu}}],\,\mathcal{D}\bb{p_{\mu}}\right] = 0.
\label{pp07}
\end{equation}\normalsize
Thus we can reobtain the set of novel generators $k_{\ii}$ and $j_{\ii}$ that satisfy the ordinary Lorentz algebra of (\ref{pp02A})
in terms of
\small\begin{equation}
M_{\mu\nu} \equiv L_{\mu\nu} + \left[L_{\mu\nu},\,\mathcal{D}\bb{p_{\mu}}\right],
\label{pp08}
\end{equation}\normalsize
by simply observing that
\small\begin{equation}
\left[L_{\mu\nu},\,\mathcal{D}\bb{\lambda, p_{\mu}}\right] = \lambda \left(p_{\mu} n_{\nu} - p_{\nu} n_{\mu}\right) p_{\a}\partial^{\a},
\label{pp09}
\end{equation}\normalsize
so that
\small\begin{equation}
j_{\ii} = \epsilon^{\ii\j\k}M_{\j\k} = J_{\ii} + 2 \lambda \epsilon^{\ii\j\k} p_{\j}n_{\k} p_{\nu}\partial^{\nu}
\label{pp10}
\end{equation}\normalsize
and
\small\begin{equation}
k_{\ii} = K_{\ii} + \lambda (p_{\ii}n_{\0}- p_{\0}n_{\ii}) p_{\nu}\partial^{\nu},
\label{pp10B}
\end{equation}\normalsize
which reproduces the same commuting rules of (\ref{pp02A}) with $J_{\ii}\mapsto j_{\ii}$ and $K_{\ii}\mapsto k_{\ii}$. 

These transformations clearly do not preserve the usual quadratic invariant in the momentum space.
But there is a modified invariant $||U\bb{D\bb{\lambda, p_{\mu}}}\circ p_{\mu}||^{\2} = M^{^{\2}}$
which leads to  the following dispersion relation,
\small\begin{equation}
||U\bb{D\bb{\lambda, p_{\mu}}}\circ p_{\mu}||^{\2} = \frac{p^{\2}}{(1 - \lambda \, p_{\nu}n^{\nu})^{\2}} = M^{^{\2}}
\label{pp11}
\end{equation}\normalsize

In spite of several particular cases of the above transformation being sufficiently explored in the literature, in opposition to the usual dilatation and special conformal transformation (transvection), the above result does not preserve the light cone.
The exception occurs when the connection between two phase-space points $k_{\mu}$ and $p_{\mu}$ satisfy the condition $p_{\nu}n^{\nu} = k_{\mu}n^{\mu}$.
Conformal transformations can convert timelike vectors into spacelike vectors, which, in principle, brings up some problems with the preservation of causality.
In fact, in quantum field theory, the observable operators associated with regions separated by spacelike distances have to commute one with each other.
Therefore, this notion of causality would be ambiguous due to the conformal symmetry.
In certain situations, such a difficulty could be overcome by the observation that quantum fields are subjected to nonlocal transformations \cite{Sch74}, which, however, does not represent a general solution.

It is also convenient to emphasize that, by embedding the Lorentz algebra into a modified conformal group, the new operators $k_{\ii}$ and $j_{\ii}$ precisely obtained from the original {\em boost} and rotation generators via the adjoint action of a $p_\mu$-dependent transformation $U\bb{D\bb{\lambda, p_{\mu}}}$ still satisfy the Lorentz algebra, but now in a nonlinear representation.
From Eq.(\ref{pp07}) the transformation $U\bb{D\bb{\lambda, p_{\mu}}}$ preserves the structure of the Lorentz algebra, and $k_i$ and $j_i$  play the role of the generators of the Lorentz algebra in the {\em preferred} frame, where the laws of physics are invariant under rotations and translations therein \cite{Col97}, but not in any other frame \cite{Col99}.
More generically, the denomination of Lorentz-invariance violation concerns some modification on the dispersion relation for a particle \cite{AmeDSR,DeG06}, and this approach is used in, e.g., \cite{AmeDSR,Col97,Bet06,kifune}.
This character is also shared by DSR, where  the conventional energy-momentum dispersion relation is generalized, but in the context of a parameter related to the Planck mass scale \cite{Gho06}.

\subsection{Spacetime coordinate formulation \label{st}}

The nonlinear action of the Lorentz transformations in momentum space has no immediate equivalent transformation in coordinate space.
Once in momentum space, one may ask how to reproduce the spacetime coordinate formulation.
The most prominent prescription is to define the spacetime coordinates as the generators of shifts in
momentum space\footnote{It is important to notice the contravariant form of the spacetime coordinate $x^{\mu} = (x^{\0}, x^{\ii})$ in order
to get the right signals of covariant representation of the subsequent results.
Here $x^{\mu} = (x^{\0}, x^{\ii}) \equiv i \tilde{\partial}^{\mu} = i (\tilde{\partial}^{\0}, - \tilde{\partial}^{\ii})$ since the
differentiation with respect to a covariant component $\tilde{p}_{\mu} = (\tilde{p}_{\0}, -\tilde{p}_{\ii})$ gives a contravariant
vector operator $\tilde{\partial}^{\mu} = (\tilde{\partial}^{\0}, - \tilde{\partial}^{\ii})$.},
\small\begin{equation}
\left.
\begin{array}{rlrlr}
t\equiv x^{\0}&\equiv& i \tilde{\partial}^{\0} &\equiv& i \frac{\partial}{\partial \tilde{p}_{\0}}\\
x^{\ii}&\equiv& - i \tilde{\partial}^{\ii} &\equiv& - i \frac{\partial}{\partial \tilde{p}_{\ii}}
\end{array}
\right\}
x^{\mu}\equiv  i \tilde{\partial}^{\mu} \equiv i \frac{\partial}{\partial \tilde{p}_{\mu}} \equiv (i \frac{\partial}{\partial \tilde{p}_{\0}}, - i \frac{\partial}{\partial \tilde{p}_{\ii}})
\label{new01}
\end{equation}\normalsize
as it was similarly proposed in \cite{Kow02}.
In this case, shifts are not pure additive constants.
For the general case of a nonlinear theory with the energy-momentum transforming as
\small\begin{equation}
\tilde{p}_{\0} = f\bb{p_{\0},p_{\ii}} \,p_{\0} ~~~~\mbox{and}~~~~\tilde{p}_{\ii} = f\bb{p_{\0},p_{\ii}}\, p_{\ii},
\label{new02}
\end{equation}\normalsize
which parameterizes the action of the transformation $U\bb{\mathcal{D}}$, the shifts in the momentum space can be described in terms of
the system of coupled equations,
\small\begin{eqnarray}
\delta \tilde{p}_{\0} &=& \frac{\partial \tilde{p}_{\0}}{\partial p_{\0}}\,\delta p_{\0} + \frac{\partial \tilde{p}_{\0}}{\partial p_{\j}}\,\delta p_{\j}\nonumber\\
\delta \tilde{p}_{\ii} &=& \frac{\partial \tilde{p}_{\ii}}{\partial p_{\0}}\,\delta p_{\0} + \frac{\partial \tilde{p}_{\ii}}{\partial p_{\j}}\,\delta p_{\j}
\label{new03}
\end{eqnarray}\normalsize
For small energy shifts ($\delta \tilde{p}_{\0} = \varepsilon$ and $\delta \tilde{p}_{\ii} = 0$)
we evaluate the above system of equations in order to obtain $t$, and for small momentum shifts
($\delta \tilde{p}_{\0} = 0$ and $\delta \tilde{p}_{\ii} = \varepsilon_{\ii}$)
we evaluate the equations in order to obtain $x_{\ii}$.
It can be summarized by rewriting the modified spacetime coordinates as
\small\begin{eqnarray}
t \equiv x^{\0} &\equiv&
~~i \left(\left.\frac{\delta p_{\0}}{\delta\tilde{p}_{\0}}\right|_{_{\delta\tilde{p}_{\ii}=0}}\frac{\partial}{\partial p_{\0}} +
\left.\frac{\delta p_{\j}}{\delta\tilde{p}_{\0}}\right|_{_{\delta\tilde{p}_{\ii}=0}}\frac{\partial}{\partial p_{\j}}\right)\nonumber\\
x^{\ii}&\equiv&
-i \left(\left.\frac{\delta p_{\0}}{\delta\tilde{p}_{\ii}}\right|_{_{\delta\tilde{p}_{\0}=0}}\frac{\partial}{\partial p_{\0}} +
\left.\frac{\delta p_{\j}}{\delta\tilde{p}_{\ii}}\right|_{_{\delta\tilde{p}_{\0}=0}}\frac{\partial}{\partial p_{\j}}\right)
\label{new04}
\end{eqnarray}\normalsize
which, from now on, can be utilized for computing the generalized commutation relations $[x^{\mu}, p_{\nu}]$
and $[x^{\mu}, x_{\nu}]$.

To illustrate the applicability of the above equations in deriving the real space coordinate transformations in agreement
with the convention specified by Eq.(\ref{new01}), let us turn back to the dilatation embedded in a LV
system where the dispersion relation is reconstructed in terms of (\ref{pp11}).
For this case, the energy and momentum components transform respectively as
\small\begin{equation}
\tilde{p}_{\0} = \frac{p_{\0}}{1 - \lambda \, p_{\nu}n^{\nu}}~~\mbox{and}~~\tilde{p}_{\ii} = \frac{p_{\ii}}{1 - \lambda \, p_{\nu}n^{\nu}},
\label{new05}
\end{equation}\normalsize
and the evolution of the system (\ref{new03}) leads to the following results;
for small energy shifts $\varepsilon$ we obtain
\small\begin{eqnarray}
\delta p_{\0} &=& (1 - \lambda \, p_{\nu}n^{\nu})(1 - \lambda \, p_{\0}n^{\0}) \varepsilon, \nonumber\\
\delta p_{\ii} &=& -\lambda \, p_{\ii} n_{\0}(1 - \lambda \, p_{\nu}n^{\nu}) \varepsilon, 
\label{new06}
\end{eqnarray}\normalsize
and the corresponding time coordinate 
\small\begin{eqnarray}
x^{\0} &=& i (1 - \lambda \, p_{\nu}n^{\nu})(\partial^{\0} - \lambda n^{\0} \,p_{\nu}\partial^{\nu});
\label{new06B}
\end{eqnarray}\normalsize
for small momentum shifts $\varepsilon_{\ii}$ we obtain
\small\begin{eqnarray}
\delta p_{\0} &=& -\lambda \, p_{\0} (1 - \lambda \, p_{\nu}n^{\nu}) n^{\ii} \varepsilon_{\ii}, \nonumber\\
\delta p_{\ii} &=&  (1 - \lambda \, p_{\nu}n^{\nu})(\varepsilon_{\ii} + \lambda \, p_{\ii} n^{\j}\varepsilon_{\j}),
\label{new07}
\end{eqnarray}\normalsize
and the corresponding space coordinates
\small\begin{eqnarray}
x^{\ii} &=& - i (1 - \lambda \, p_{\nu}n^{\nu})(\partial^{\ii} + \lambda n^{\ii} \,p_{\nu}\partial^{\nu}).
\label{new07B}
\end{eqnarray}\normalsize
Hence the spacetime coordinate in the contravariant notation is   
\small\begin{eqnarray}
x^{\mu} = (x^{\0}, x^{\ii}) &=& i (1 - \lambda \, p_{\nu}n^{\nu})(\partial^{\mu} - \lambda n^{\mu} \,p_{\nu}\partial^{\nu}),
\label{new08}
\end{eqnarray}\normalsize
as noticed in \cite{Mag03}, very similar to the Snyder's noncommutative geometry \cite{Sny47},
\small\begin{eqnarray}
x^{\mu} &=& i (\partial^{\mu} - \lambda p^{\mu} \,p_{\nu}\partial^{\nu}).
\label{new09}
\end{eqnarray}\normalsize
We can easily check that the spacetime coordinates all commute with each other and that the commutators of spacetime
coordinates with momentum coordinates retract a novel energy and direction dependent Planck's constant,
\small\begin{equation}
[x^{\mu}, x_{\nu}] = 0 ~~\mbox{and}~~ [x^{\mu}, p_{\nu}] = (1 - \lambda \, p_{\nu}n^{\nu})(\delta^{\mu}_{\nu} - \lambda\, n^{\mu}p_{\nu}).
\label{new10}
\end{equation}\normalsize
In some sense, it reflects the suppression of the uncertainty principle for circumstances where $\lambda \, p_{\nu}n^{\nu} \approx 1$.
In addition, from a particular point of view, we are also reproducing the result obtained in \cite{Mag03} since, as we shall describe in the
following, it corresponds to the above results with a particular choice of $n_{\mu}$.

\subsection{LV systems with an invariant preferred direction} %

The generalized nonlinear representation of the Lorentz group in the momentum space is given by
\small\begin{equation}
\tilde{\Lambda}_{\mu}^{\nu} = U^{^{\mi 1}}\hspace{-0.35 cm}\bb{\mathcal{D}}\, \Lambda_{\mu}^{\nu}\, U\bb{\mathcal{D}}.
\label{new12}
\end{equation}\normalsize
	
In evaluating this expression, we note that $\mathcal{D}\bb{\lambda, p_{\mu}}$ acts on everything to the right, and $p_{\mu}$
means the vector dependence of the function-variable immediately to the right.
The general rule can be illustrated by the transformation
\small\begin{equation}
\left.
\begin{array}{l}
U\bb{\mathcal{D}\bb{\lambda, p_{\mu}}} \circ p_{\mu} = \tilde{p}_{\mu}\bb{\lambda, p_{\mu}}\\
U\bb{\mathcal{D}\bb{\lambda, p^{\prime}_{\mu}}} \circ p^{\prime}_{\mu} = \tilde{p}^{\prime}_{\mu}\bb{\lambda, p^{\prime}_{\mu}}
\end{array}
\right\} \Rightarrow 
\tilde{p}^{\prime}_{\mu}\bb{\lambda, p^{\prime}_{\mu}} = \Lambda_{\mu}^{\nu} \tilde{p}_{\nu}
\label{new13}
\end{equation}\normalsize
which leads to
\small\begin{equation}
p^{\prime}_{\mu}\bb{\lambda, p^{\prime}_{\mu}} \equiv 
U^{^{\mi 1}}\hspace{-0.35 cm}\bb{\mathcal{D}}\circ \tilde{p}^{\prime}_{\mu} =
U^{^{\mi 1}}\hspace{-0.35 cm}\bb{\mathcal{D}}\, \Lambda_{\mu}^{\nu}\, (U\bb{\mathcal{D}} \circ p_{\nu})=
\tilde{\Lambda}_{\mu}^{\nu} p_{\nu}
\label{new14}
\end{equation}\normalsize

Using this rules for $\mathcal{D}\bb{\lambda, p_{\mu}}$ of the Eq.~(\ref{pp03}) that leads to the relations of (\ref{new05}),
one finds that the boosts are now given by
\small\begin{eqnarray}
p^{\prime}_{\0}&=& \varrho\bb{p_{\0},p_{\parallel},p_{\perp}} \,\gamma \left[p_{\0} - \beta p_{\parallel}\right]\nonumber\\
p^{\prime}_{\parallel}&=&\varrho\bb{p_{\0},p_{\parallel},p_{\perp}} \,\gamma \left[p_{\parallel} - \beta p_{\0}\right]\nonumber\\
p^{\prime}_{\perp}&=&\varrho\bb{p_{\0},p_{\parallel},p_{\perp}} \,p_{\perp}
\label{new15}
\end{eqnarray}\normalsize
with
\small\begin{eqnarray}
\varrho\bb{p_{\0},p_{\parallel},p_{\perp}}&=&\left\{ 1 + \lambda(\gamma - 1) \,p_{\mu}n^{\nu}\right.\nonumber\\
&&~~~\left.- \lambda \gamma \mbox{\boldmath$\beta$} \cdot \left( \mbox{\boldmath$p$} n_{\0} -  \mbox{\boldmath$n$} p_{\0}\right) - \lambda \mbox{\boldmath$n$}\cdot \lambda \mbox{\boldmath$p$}_{\perp}\right\}^{\mi \1}~~~
\label{new16}
\end{eqnarray}\normalsize
where the symbols $\parallel$ and $\perp$ respectively stand for the momentum components parallel and perpendicular to the boost ({\boldmath$\beta$}).
The relations in (\ref{new15}) reduce to the usual transformations when $\lambda \, p_{\nu}n^{\nu} << 1$.

\subsection{LV systems with an invariant energy scale}

All the previous results can be specialized by setting the invariant preferred direction 
$n_{\mu}$ as a spacelike vector given by $n_{\mu} = (1,\,0,\,0,\,0)$.
In this case, we can immediately reproduce the results of a LV system with an invariant energy scale introduced in \cite{Mag02}.
In fact, with $n_{\mu} = (1,\,0,\,0,\,0)$, the modified Lorentz transformations described by the Eq.~(\ref{new15})
will be written exactly in the same form, just with the modified coefficient 
\small\begin{eqnarray}
\varrho\bb{p_{\0},p_{\parallel},p_{\perp}}&=&\left\{ 1 + \lambda(\gamma - 1) \,p_{\0}-\lambda \gamma \beta p_{\parallel} \right\}^{\mi \1}
\label{new16C}
\end{eqnarray}\normalsize
In this case, when $p_{\0} =1/ \lambda$, we can easily verify that $p^{\prime}_{\0}  = p_{\0} = 1/ \lambda$, i. e. at the Planck energy scale, the energy component transformation illustrate an
invariant energy scale, which was extensively studied in \cite{Mag02,Mag03}.
In an analogous way, the spacetime coordinate formulation will be subtly and conveniently modified.
For small energy shifts $\varepsilon$ we obtain the corresponding time coordinate 
\small\begin{eqnarray}
x^{\0} &=& i (1 - \lambda \, p_{\0})(\partial^{\0} - \lambda \,p_{\nu}\partial^{\nu}),
\label{new06BC}
\end{eqnarray}\normalsize
and for small momentum shifts $\varepsilon_{\ii}$ we obtain the corresponding space coordinates
\small\begin{eqnarray}
x^{\ii} &=& - i (1 - \lambda \, p_{\0})\partial^{\ii}.
\label{new07BC}
\end{eqnarray}\normalsize
Differently from the general case, now the spacetime coordinate transformation 
are resumed by the noncovariant form
\small\begin{eqnarray}
x^{\mu} &=& i (1 - \lambda \, p_{\0})(\partial^{\mu} - \lambda \delta^{\mu\0}\,p_{\nu}\partial^{\nu}),
\label{new08C}
\end{eqnarray}\normalsize
so that each commutator has to be written separately in a noncovariant way as
\small\begin{eqnarray}
\left[x^{\mu}, x_{\nu}\right] &=& 0, \nonumber\\
\left[x^{\ii}, p_{\j}\right] &=& \delta^{\ii}_{\j} (1 - \lambda \, p_{\0}),\nonumber\\
\left[x^{\0}, p_{\ii}\right] &=& -\lambda (1 - \lambda \, p_{\0}) p_{\ii},\nonumber\\
\left[x^{\0}, p_{\0}\right] &=& (1 - \lambda \, p_{\0})^{\2}.
\label{new10C}
\end{eqnarray}\normalsize

In agreement with the above results, considerations based on quantum gravity and black hole physics strongly indicate that,
at the smallest scale, spacetime coordinates become, for instance, noncommutative \cite{Bet06,Sny47,Ban06}.
In this sense, further implications of all of those scenarios are being currently investigated.

\section{Free propagating Dirac particle equation of motion correlated to VSR}

The formalism of VSR has been expanded for studying some peculiar aspects of neutrino physics with the VSR subgroup chosen to be the 4-parameter group SIM(2) \cite{Gla06B}.
Since neutrinos are now known to be massive, several mechanisms have been contrived to remedy the absence of neutrino mass in the Standard Model Lagrangian \cite{Bil03}.
The framework of VSR admits the unconventional possibility of neutrino masses that neither violate lepton number nor require additional sterile states \cite{Gla06B}.
In the {\em Dirac} picture, lepton number is conserved with neutrinos acquiring mass via Yukawa couplings to sterile SU(2)-singlet neutrinos \cite{Zub98,Kim93}.
In the {\em Majorana} picture, lepton number is violated and neutrino masses result from a seesaw mechanism involving heavy sterile states or via dimension-6 operators resulting from unspecified new interactions \cite{Gel79,Moh86}. 
In spite of not being Lorentz invariant, the lepton number conserving neutrino masses are VSR invariant.
There is no guarantee that neutrino masses have a VSR origin, but if so their sizes may be an indication of the magnitude of LV effects in other sectors, for instance, as a suggestion to the examination of the existence of a preferred axis in the cosmic radiation anisotropy \cite{Mag05}.

In this section we intend to show how to adequate the results of VSR to a Lorentz-invariance violation system reconstructed by means of modified conformal transformations acting on the Lorentz generators.
As we shall see in the following, one way to do this is to combine each boost/rotation with an specific transformation in which we introduce a preferential direction with the aid of a lightlike vector defined as $n_{\mu}$($\equiv(1,0,0,1)$), $n^{2}=0$. 

\subsection{The equation of motion} 

The transformation has to be chosen as to bring an equation of motion which recovers the dynamics of the equation introduced in \cite{Gla06B},
\small\begin{equation}
\left(\gamma^{\mu}p_{\mu} - \frac{m^{\2}_{\nu}}{2}\frac{\gamma^{\mu}n_{\mu}}{p_{\lambda}n^{\lambda}}\right)(1-\gamma^{\5})\nu\bb{x} = 0,
\label{app01}
\end{equation}\normalsize
which admits the quoted unconventional possibility of neutrino masses that neither violate lepton number nor require additional sterile states.
The above dynamics requires only two degrees of freedom for a particle carrying lepton number: one for the neutrino with positive lepton number, and one for the antineutrino with negative lepton number.
The VSR neutrino at rest is necessarily an eigenstate of angular momentum in the preferred direction with eigenvalue $+1/2$.
Thus, stimulated by the ideas of Cohen and Glashow \cite{Gla06A,Gla06B}, we search for convenient unitary transformations $U$ acting on the usual Lorentz generators in order to recover the equation of motion for a free propagating fermionic particle.
We expect that, in a relatively high energy scale, the corresponding equation will be reduced to the Glashow Eq.(\ref{app01}), and, in a relatively low energy scale, it will be reduced to the usual Dirac equation for a free propagating fermionic particle. 

Let us recover the standard definitions for the momentum space and the corresponding algebra given by the Eqs.~(\ref{pp02}-\ref{pp02A}). 
In order to introduce the nonlinear action that modifies the ordinary Lorentz generators but, however,
preserves the structure of the algebra, we suggest the following {\em ansatz} for a generalized transformation, 
\small\begin{equation}
D \equiv (a\bb{y}\,p_{\mu} + b\bb{y}\,n_{\mu})\partial^{\mu}
\label{app03}
\end{equation}\normalsize
which acts on the momentum space as 
\small\begin{equation}
D \circ p_{\mu} \equiv a\bb{y}\,p_{\mu} + b\bb{y}\,n_{\mu}
\label{app04}
\end{equation}\normalsize
where $y = p_{\nu}n^{\nu}$.
For such a definition, we have
\small\begin{equation}
\partial_{\mu} f\bb{y} = n_{\mu}\, \frac{\mbox{d}f\bb{y}}{\mbox{d}y} \equiv n_{\mu} f^{\prime}\bb{y}, ~~~~f\bb{y} = a\bb{y}, \, b\bb{y},
\label{app05}
\end{equation}\normalsize
and, consequently, $p^{\mu}\partial_{\mu} f\bb{y}= y\, f^{\prime}\bb{y}$ and $n^{\mu}\partial_{\mu} f\bb{y}= 0$. 
We assume the new action can be considered to be a nonstandard and nonlinear embedding of the Lorentz group into a modified conformal group which, as we shall notice in the following for the case of main interest, despite the modifications, satisfies precisely the ordinary Lorentz algebra (\ref{pp02A}).
To exponentiate the new action, we note that 
\small\begin{equation}
k^{\ii} = U^{^{\mi 1}}\hspace{-0.35 cm}\bb{D}\, K^{\ii}\, U\bb{D} ~~\mbox{and}
~~ j^{\ii} = U^{^{\mi 1}}\hspace{-0.35 cm}\bb{D} \,J^{\ii} \,U\bb{D}
\label{app06}
\end{equation}\normalsize
where the $y$-dependent transformation $U\bb{D}$ is given by $U\bb{D\bb{y}} \equiv{\exp[D\bb{y}]}$.
The nonlinear representation is then generated by $U\bb{D\bb{y}} \circ p_{\mu}$ and, despite not being unitary ($U\bb{D\bb{y}} \circ p_{\mu} \neq p_{\mu}$), it has to preserve the structure of the algebra.
Thus, when we assume 
\small\begin{equation}
\left[[L_{\mu\nu},\,D\bb{y}],\,D\bb{y}\right] = 0
\label{app07}
\end{equation}\normalsize
we can reobtain a set of generators (in terms of $k_{\ii}$ and $j_{i}$) which satisfy the ordinary Lorentz algebra of (\ref{pp02A}).  
At this point, in order to explicitly obtain the operator $D\bb{y}$ which satisfies the relation (\ref{app07}), we firstly compute the commutation relation
\small\begin{equation}
[L_{\mu\nu},\,D\bb{y}] = \kappa_{\mu\nu}(a^{\prime}\bb{y}\,p_{\alpha} + b^{\prime}\bb{y}\,n_{\alpha})\partial^{\alpha}
+ b\bb{y} \, d_{\mu\nu},
\label{app08}
\end{equation}\normalsize
for which we have defined the parameters
\small\begin{eqnarray}
&&\kappa_{\mu\nu} = p_{\mu} n_{\nu} - p_{\nu} n_{\mu}\nonumber\\
&&d_{\mu\nu} = n_{\nu} \partial_{\mu} - n_{\mu} \partial_{\nu}.
\label{app09}
\end{eqnarray}\normalsize
From the above definitions we obtain the useful relations
\small\begin{eqnarray}
&&D \,\kappa_{\mu\nu} = a\bb{y} \kappa_{\mu\nu}\nonumber\\
&&d_{\mu\nu} \,D = a\bb{y} d_{\mu\nu}\nonumber\\
&&D \,d_{\mu\nu} = 0 
\label{app10}
\end{eqnarray}\normalsize
which are essential in computing $[[L_{\mu\nu},\,D\bb{y}],\,D\bb{y}]$.
The first part of the r.h.s. of the Eq.(\ref{app08}) then leads to the commutation relation
\small\begin{eqnarray}
\lefteqn{[\kappa_{\mu\nu}(a^{\prime}\bb{y}\,p_{\alpha} + b^{\prime}\bb{y}\,n_{\alpha})\partial^{\alpha},\,D\bb{y}] =}\nonumber\\
&& \kappa_{\mu\nu}\left\{\left[y (a^{^{\prime} \2}\bb{y} - a\bb{y}\, a^{\prime\prime}\bb{y}) - a\bb{y}\,a^{\prime}\bb{y}\right] p^{\alpha}\partial_{\alpha}\right.\nonumber\\ 
&&~~~~+ \left.\left[a^{\prime}\bb{y}(y\, b^{\prime}\bb{y} - b\bb{y}) -y\, a\bb{y}\,b^{\prime\prime}\bb{y}\right] n^{\alpha}\partial_{\alpha}\right\},
\label{app11}
\end{eqnarray}\normalsize
and the second part gives
\small\begin{equation}
[b\bb{y} \, d_{\mu\nu},\,D\bb{y}] = a\bb{y} (b\bb{y} - y \,b^{\prime}\bb{y}) d_{\mu\nu}.
\label{app12}
\end{equation}\normalsize
In order to satisfy the condition for preserving the Lorentz algebra (\ref{app07}), the $y$-dependent coefficients can be obtained by evaluating the coupled ordinary differential equations: 
\small\begin{eqnarray}
y (a^{^{\prime} \2}\bb{y} - a\bb{y}\, a^{\prime\prime}\bb{y}) - a\bb{y}\,a^{\prime}\bb{y} = 0 &&~~~~(a.1)\nonumber\\
a^{\prime}\bb{y}(y\, b^{\prime}\bb{y} - b\bb{y}) -y\, a\bb{y}\,b^{\prime\prime}\bb{y}  = 0 &&~~~~(a.2)\nonumber\\
a\bb{y} (b\bb{y} - y \,b^{\prime}\bb{y})                                                = 0 &&~~~~(a.3)\nonumber
\end{eqnarray}\normalsize
for which we have two types of solutions: 

\paragraph*{Type-I}  $a\bb{y}=0$ and $\forall ~b\bb{y}$.

\paragraph*{Type-II}  $a\bb{y}= A y^{\n}, n\in \mathcal{R}$ and $b\bb{y}= B y$ where $A$ and $B$
are constants with respective dimensions given by $[[A]]\equiv m^{\mi\n}$ and $[[B]]\equiv m^{\mi \1}$.

Other choices for $U\bb{D}$($\equiv U\bb{y, p^{\2}, m^{\2}}$) are possible and lead to different boost generators,
but the proposed Type-I solution provides the simplest analytical procedure for recovering the Eq.~(\ref{app01}).
For a Type-I solution where $a\bb{y} = 0$ and
\small\begin{equation}
b\bb{y} = \frac{\alpha \,m^{\2}}{1- 2 \alpha y},
\label{app15}
\end{equation}\normalsize
we can easily verify that
\small\begin{eqnarray}
\lefteqn{D\bb{y} \equiv \frac{\alpha\,m^{\2}}{1- 2 \alpha y}\,n_{\mu}\partial^{\mu}}\nonumber\\
&\Rightarrow&D\bb{y}\circ p_{\mu}\equiv \frac{\alpha\,m^{\2}}{1- 2 \alpha y}\,n_{\mu}  \nonumber\\
&\Rightarrow&U\bb{D\bb{y}}\circ p_{\mu}\equiv p_{\mu} + \frac{\alpha\,m^{\2}}{1- 2 \alpha y}\,n_{\mu}.\nonumber\\  
\label{app16}
\end{eqnarray}\normalsize
At the same time, we obtain the new generators $j_{\ii}$ and $k_{\ii}$ from Eq.~(\ref{app06}),
\small\begin{eqnarray}
j_{\1}&=&J_{\1}+ 2p_{\2}\,m^{\mi\2}\,b^{^{\2}}\hspace{-0.15cm}\bb{y}\,n_{\lambda}\partial^{\lambda} + b\bb{y}\, \partial_{\2}\nonumber\\
j_{\2}&=&J_{\2}- 2p_{\1}\,m^{\mi\2}\,b^{^{\2}}\hspace{-0.15cm}\bb{y}\,n_{\lambda}\partial^{\lambda} - b\bb{y}\, \partial_{\1}\nonumber\\
j_{\3}&=&J_{\3}\nonumber\\
k_{\1}&=&K_{\1}+ 2p_{\1}\,m^{\mi\2}\,b^{^{\2}}\hspace{-0.15cm}\bb{y}\,n_{\lambda}\partial^{\lambda} + b\bb{y} \,\partial_{\1}\nonumber\\
k_{\2}&=&K_{\2}+ 2p_{\2}\,m^{\mi\2}\,b^{^{\2}}\hspace{-0.15cm}\bb{y}\,n_{\lambda}\partial^{\lambda} + b\bb{y}\, \partial_{\2}\nonumber\\
k_{\3}&=&K_{\3}+ 2(p_{\3}-p_{\0})m^{\mi\2}\, b^{^{\2}}\hspace{-0.15cm}\bb{y}\,n_{\lambda}\partial^{\lambda} + b\bb{y}\,(\partial_{\3}-\partial_{\0})
\label{app16B}
\end{eqnarray}\normalsize
from which we can reconstruct the Lorentz algebra as
\small\begin{equation}
[R^{\ii}, T^{\j}] = \epsilon^{\ii\j\k}T_{\k};~~~~[R^{\ii}, R^{\j}] = [T^{\ii}, T^{\j}] = \epsilon^{\ii\j\k}R_{\k}
\label{app16C}
\end{equation}\normalsize
with
\small\begin{eqnarray}
R_{\1(\2)}&=& \frac{1}{\sqrt{2}} (k_{\1(\2)}+\bb{-} j_{\2(\1)})\nonumber\\
T_{\1(\2)}&=& \frac{1}{\sqrt{2}} (j_{\1(\2)}+\bb{-} k_{\2(\1)})\nonumber\\
R_{\3}&=& j_{\3}\nonumber\\
T_{\3}&=& k_{\3}
\label{app16D}
\end{eqnarray}\normalsize
These transformations clearly do not preserve the usual quadratic invariant in the momentum space.
But there is a modified invariant $||U\bb{D\bb{y}}\circ p_{\mu}||^{\2} = M^{^{\2}}\hspace{-0.15cm}\bb{\alpha}$ which leads to
the following dispersion relation,
\small\begin{eqnarray}
||U\bb{D\bb{y}}\circ p_{\mu}||^{\2} = p^{\2} + \frac{2\, y\,m^{\2}\,\alpha}{1- 2 \alpha y} =M^{^{\2}}\hspace{-0.15cm}\bb{\alpha}
\label{app17}
\end{eqnarray}\normalsize
Imposing the constraint $p^{\2}= m^{\2}$, which is also required by the VSR theory, we have the Casimir invariant 
\small\begin{equation}
M^{^{\2}}\hspace{-0.15cm}\bb{\alpha} = \frac{m^{\2}}{1- 2 \alpha y}.
\label{app18}
\end{equation}\normalsize
for which the $U$-invariance can be easily verified when we apply the transformation $U\bb{D\bb{y}}$.
The fact that the algebra of the symmetry group remains the same suggests that perhaps the standard spin connection formulation of relativity is still valid.
In this sense, the above dispersion relation can also be obtained from the dynamic equation for a fermionic particle,
\small\begin{eqnarray}
\lefteqn{\left[\gamma^{\mu}\left(U\bb{D\bb{y}}\circ p_{\mu}\right) - M\bb{\alpha}\right]\nu_{_{\L}}\bb{x} = 0} \nonumber\\
&&\Rightarrow\left(\gamma^{\mu}p_{\mu} + \frac{m^{\2} \alpha}{1 - 2\,\alpha\,y }\gamma^{\mu}n_{\mu} - M\bb{\alpha} \right)\nu_{_{\L}}\bb{x} = 0.
\label{app19}
\end{eqnarray}\normalsize
Alternatively, as pointed out in \cite{Mag03}, for a comparative purpose to all these classes of LV models,
dispersion relations may be derived from calculations in a theory such as loop quantum gravity \cite{Gam99}.

By setting $[[m]]^{\mi\1}$ values to $\alpha$, for instance, $\alpha = \pm 1/m,\,\pm m/E^{\2}_{\P\l}$
(where $E_{\P\l}$ is the Planck's energy),
we are able to analyze the low and the high energy limits.
In the high energy limit where $\alpha y >> 1$, the Eq.~(\ref{app19}) is reduced to
\small\begin{equation}
\left(\gamma^{\mu}p_{\mu} - \frac{m^{\2}}{2\,y }\gamma^{\mu}n_{\mu} - M\bb{\alpha}\right)\nu_{_{\L}}\bb{x} = 0.
\label{app19A}
\end{equation}\normalsize
and since $M^{^{\2}}\hspace{-0.15cm}\bb{\alpha}\approx \frac{m^{\2}}{2\,|\alpha\,y|} << m^{\2}$,
in spite of not being necessary, we can eliminate the dependence on $\alpha$ since 
the $M\bb{\alpha}$ term becomes irrelevant in the above equation.
Thus we recover the {\em VSR Cohen-Glashow} equation (\ref{app01})\cite{Gla06B}
and its corresponding dispersion relation, as we have proposed from the initial part of this letter.

In the low energy limit where $|\alpha y| << 1$, the Eq.~(\ref{app19}) is reduced to
\small\begin{equation}
\left(\gamma^{\mu}p_{\mu} + m^{\2}\,\alpha \gamma^{\mu}n_{\mu} - m \right)\nu_{_{\L}}\bb{x} = 0,
\label{app19B}
\end{equation}\normalsize
whose the quadratic form is
\small\begin{equation}
\left(p^{\2} + 2 m^{\2}\,\alpha y - m^{\2}\right)\nu_{_{\L}}\bb{x} \approx  \left(p^{\2} -  m^{\2}\right)\nu_{_{\L}}\bb{x} = 0,
\label{app19C}
\end{equation}\normalsize
i.e. when $|\alpha y| << 1$ the effective contribution from the second term of Eq.~(\ref{app19B}) is negligible and the equation can be reduced to the usual (low energy limit) {\em Dirac} equation for a free propagating particle.
Such an important result could also be reproduced in a more direct way if we initially assumed
a natural energy scale where $|m \alpha| << 1$, for instance, when $\alpha =  m/E^{\2}_{\P\l}$.
Since $M$ is reduced to $m$, the Eq.~(\ref{app19B}) is immediately reduced to the {\em Dirac} equation.
For all the above analysis, the dispersion relation $p^{\2}= m^{\2}$ is maintained and the considerations
about the preservation of the light cone are analogous to the ones presented in the previous section.

\subsection{Real space formulation}  

By assuming the spacetime formulation prescribed in the section \ref{st},
we can compute the spacetime coordinates as the generators of shifts in momentum space.
Now the energy and momentum components transform respectively as
\small\begin{equation}
\tilde{p}_{\0} = 
p_{\0} + \frac{m^{\2}\, \alpha}{1 - 2\,\alpha \, p_{\nu}n^{\nu}}n_{\0}
~~\mbox{and}~~
\tilde{p}_{\ii} = p_{\ii} + \frac{m^{\2}\, \alpha}{1 - 2\,\alpha \, p_{\nu}n^{\nu}}n_{\ii}
\label{new05D}
\end{equation}\normalsize
and the evolution of the system (\ref{new03}) leads to the following results.
For small energy shifts $\varepsilon$ we obtain
\small\begin{eqnarray}
\delta p_{\0} &=& \left[1 - \frac{2\,m^{\2}\, \alpha^{\2} n_{\0}^{\2}}{\left(1 - 2\,\alpha \, p_{\nu}n^{\nu}\right)^{\2}}\right] \varepsilon, \nonumber\\
\delta p_{\ii} &=& - \frac{2\,m^{\2}\, \alpha^{\2} n_{\0}}{\left(1 - 2\,\alpha \, p_{\nu}n^{\nu}\right)^{\2}}n_{\ii} \varepsilon, 
\label{new06D}
\end{eqnarray}\normalsize
and the corresponding time coordinate 
\small\begin{eqnarray}
x^{\0} &=& i 
\left(\partial^{\0} - \frac{2\,m^{\2}\, \alpha^{\2}}{\left(1 - 2\,\alpha \, p_{\nu}n^{\nu}\right)^{\2}}\,n^{\0} \,n_{\nu}\partial^{\nu}\right).
\label{new06BD}
\end{eqnarray}\normalsize
For small momentum shifts $\varepsilon_{\ii}$ we obtain
\small\begin{eqnarray}
\delta p_{\0} &=& - \frac{2\,m^{\2}\, \alpha^{\2} n_{\0}}{\left(1 - 2\,\alpha \, p_{\nu}n^{\nu}\right)^{\2}}n^{\j} \varepsilon_{\j}, \nonumber\\
\delta p_{\ii} &=&  \varepsilon_{\ii} + \frac{2\,m^{\2}\, \alpha^{\2}}{\left(1 - 2\,\alpha \, p_{\nu}n^{\nu}\right)^{\2}}(n^{\j} \varepsilon_{\j}) n_{\ii}
\label{new07D}
\end{eqnarray}\normalsize
and the corresponding space coordinates
\small\begin{eqnarray}
x^{\ii} &=& - i \left(\partial^{\ii} + \frac{2\,m^{\2}\, \alpha^{\2}}{\left(1 - 2\,\alpha \, p_{\nu}n^{\nu}\right)^{\2}}\,n^{\ii} \,n_{\nu}\partial^{\nu}\right).
\label{new07BD}
\end{eqnarray}\normalsize
Hence the spacetime coordinate in the contravariant notation is   
\small\begin{eqnarray}
x^{\mu} = (x^{\0}, x^{\ii}) &=& i \left(\partial^{\mu} - \frac{2\,m^{\2}\, \alpha^{\2}}{\left(1 - 2\,\alpha \, p_{\nu}n^{\nu}\right)^{\2}}\,n^{\mu} \,n_{\nu}\partial^{\nu}\right).
\label{new08D}
\end{eqnarray}\normalsize
We can easily check that the spacetime coordinates all commute with each other and that the covariant commutators of spacetime
coordinates with momentum coordinates retract a novel energy and direction dependent Planck's constant, $[x^{\mu}, x_{\nu}] = 0$ and
\small\begin{equation}
[x^{\mu}, p_{\nu}] = i \left(\delta^{\mu}_{\nu} - \frac{2\,m^{\2}\, \alpha^{\2}}{\left(1 - 2\,\alpha \, p_{\sigma}n^{\sigma}\right)^{\2}}\,n^{\mu}n_{\nu}\right).
\label{new10D}
\end{equation}\normalsize

The transformations presented here act on the uncertainties of position and momentum measurements made by observers.
Observers or frames of references are characterized not only by the origins of their space and time coordinates, the direction of their axis and relative velocities but also by the existence of a preferred direction.
Precision of measurements is, consequently, embodied in geometry/algebra and the laws of nature must be invariant under precision scale transformations, which deserves a more careful investigation.

\subsection{Other Lorentz algebra-preserving solutions} 

By constructing $U\bb{D\bb{y}}\equiv{\exp[D\bb{y}]}$ in terms of Type-II solutions, we can find many nonlinear realizations of the action of the Lorentz group which eventually can deserve a more careful analysis.
This leads to other forms for the modified invariants and, hence, to different dispersion relations.
We turn back to them in order to obtain generalized analytical expressions for $U\bb{D\bb{y}} \circ p_{\mu}$.
In particular, for Type-II solutions with $n = +1,\,-1, \, -1/2$, we obtain
\small
\small\begin{eqnarray}
\lefteqn{D\bb{y} \equiv (A \, y \,p_{\mu} + B y\,n_{\mu})\partial^{\mu}}\nonumber\\
&\Rightarrow& D\bb{y}\circ p_{\mu}\equiv A \, y \,p_{\mu} + B y\,n_{\mu}  \nonumber\\
&\Rightarrow& U\bb{D\bb{y}}\circ p_{\mu}\equiv \frac{p_{\mu}}{1 - A\, y} + \frac{B \,y\, n_\mu}{1 - A\, y}\nonumber\\  
&& \mbox{for}~ n = +1,~~~~~~~~~~~~~~~~~~~~~~~~~~~~~~~~~~~~~~~~~
\label{app20A}
\end{eqnarray}\normalsize
\small\begin{eqnarray}
\lefteqn{D\bb{y} \equiv (\frac{A}{y} p_{\mu} + B y\,n_{\mu})\partial^{\mu}}\nonumber\\
&\Rightarrow& D\bb{y}\circ p_{\mu}\equiv \frac{A}{y}p_{\mu} + B y\,n_{\mu}  \nonumber\\
&\Rightarrow& U\bb{D\bb{y}}\circ p_{\mu}\equiv p_{\mu}\left(1+ \frac{A}{y}\right) + B (A + y)n_{\mu}\nonumber\\  
&& \mbox{for}~ n = -1~~~~~~~~~~~~~~~~~~~~~~~~~~~~~~~~~~~~~~~~~
\label{app20B}
\end{eqnarray}\normalsize
\small\begin{eqnarray}
\lefteqn{D\bb{y} \equiv (\frac{A}{\sqrt{y}}p_{\mu} + B y\,n_{\mu})\partial^{\mu}}\nonumber\\
&\Rightarrow& D\bb{y}\circ p_{\mu}\equiv \frac{A}{\sqrt{y}}p_{\mu} + B y\,n_{\mu}  \nonumber\\
&\Rightarrow& U\bb{D\bb{y}}\circ p_{\mu}\equiv\left(1+ \frac{A}{2\sqrt{y}}\right)^{^{\2}} p_{\mu} + B \left(\sqrt{y}+ \frac{A}{2}\right)^{^{\2}}n_{\mu}\nonumber\\  
&& \mbox{for}~ n = -1/2. ~~~~~~~~~~~~~~~~~~~~~~~~~~~~~~~~~~~~~~~~~
\label{app20C}
\end{eqnarray}\normalsize
\normalsize

All the nonlinear representations here obtained correspond to particular cases of a more general prescription presented in \cite{Jud03}.
We could, for instance, conveniently take the first of them ($n = +1$) in order to obtain a novel
equation of motion with some potentiality for phenomenological studies. 
By assuming that $A = -\alpha$ and $B = - \alpha^{\2}\,m^{\2}/2 $ we would have
\small\begin{eqnarray}
\lefteqn{D\bb{y} \equiv y [\alpha\,p_{\mu} -  (\alpha^{\2}\,m^{\2}/2)\,n_{\mu}]\partial^{\mu}}\nonumber\\
&\Rightarrow& U\bb{D\bb{y}}\circ p_{\mu}\equiv \frac{p_{\mu}}{1 + \alpha\, y} - \frac{\alpha^{\2}\,m^{\2} \,y\, n_\mu}{2(1 +\alpha \, y)}  
\label{app200}
\end{eqnarray}\normalsize
Again these transformations do not preserve the usual quadratic invariant in the momentum space.
By suggesting the same connection with the spinorial formulation,
the dispersion relation related to the above transformation,
\small\begin{eqnarray}
||U\bb{D\bb{y}}\circ p_{\mu}||^{\2} =M^{^{\2}}\bb{\alpha y} =\frac{p^{\2} - \alpha^{\2}\,m^{\2}\,y^{\2}}{(1 + \alpha\,y)^{\2}},
\label{app22}
\end{eqnarray}\normalsize
can also be obtained from the dynamic equation for a fermionic particle,
\small\begin{eqnarray}
\lefteqn{\left[\gamma^{\mu}\left(U\bb{D\bb{y}}\circ p_{\mu}\right) - M\bb{\alpha}\right]\nu_{_{\L}}\bb{x} = 0} \nonumber\\
&&\Rightarrow\left(\frac{\gamma^{\mu}p_{\mu}}{1 + \alpha\,y} - \frac{\alpha^{\2}\, m^{\2}\, y\, \gamma^{\mu}n_{\mu}}{2(1 + \alpha\,y)} - M\bb{\alpha}\right)\nu_{_{\L}}\bb{x} = 0,~~
\label{app21}
\end{eqnarray}\normalsize
In the high energy limit where $\alpha y \approx 1$, the Eq.~(\ref{app21}) is reduced to
\small\begin{equation}
\left(\gamma^{\mu}p_{\mu} - \frac{\alpha m^{\2}}{2}\,\gamma^{\mu}n_{\mu} - 2 M\bb{1}\right)\nu_{_{\L}}\bb{x} = 0.
\label{app23}
\end{equation}\normalsize
which has some similitude with the Eq.~(\ref{app01}).
In the low energy limit where $\alpha y << 1$, the Eq.~(\ref{app19}) is reduced to
\small\begin{equation}
\left(\gamma^{\mu}p_{\mu} -  M\bb{0} \right)\nu_{_{\L}}\bb{x} = 0,
\label{app24}
\end{equation}\normalsize
similar to the Dirac equation.
However, for all the above solutions, it is difficult (and sometimes impossible) to constraint $M^{^{\2}}\bb{\alpha y}$ to
preserve the standard dispersion relation $p^{\2}= m^{\2}$.
This objection makes the above solutions not so interesting as the Type-I solution.
We intend to call upon these results in a subsequent work or works where we embed our discussion in a general formalism of conformal transformations
and investigate the LV effects related to the modified dynamics introduced by them.

\section{Conclusions}

In this paper we have explored some extensions of the general method for implementing nonlinear actions of the Lorentz group in the momentum space by assuming the preservation of the Lorentz algebra.
In two different approaches, we have examined some previous theoretical claims for a preferred axis in the framework of Lorentz invariance violation by generalizing the procedure for obtaining the modified Lorentz algebra, the dispersion relations, the equation of motion, the spacetime coordinates and the corresponding covariant quantum commutation relations.
First of all, we have briefly discussed the role of dilatation transformations in physics in order to introduce a deformed dilatation transformation by implementing a preferred axis $n_{\mu}$. 
In spite of the fact that our formulation represents the covariant way of performing the calculations, after analyzing the general case, we have noticed that the choice of a spacelike preferred direction with $n_{\mu} \equiv(1,0,0,1)$ 
recovers the results of the formulation of generalized Lorentz invariance with an invariant energy scale expressed in \cite{Mag02,Mag03}.
In parallel, using a generalized procedure for obtaining the dispersion relation and the equation of motion for a propagating fermionic particle, we have reexamined the case of lightlike vector $n_{\mu}$($\equiv(1,0,0,1)$), $n^{2}=0$ embedded in the framework of very special relativity (VSR).
We have shown that, in a relatively high energy scale, the corresponding equation of motion is reduced to a conserving lepton number chiral equation previously predicted in the literature \cite{Gla06B}, and, in a relatively low energy scale, it is reduced to the usual Dirac equation for a free propagating fermionic particle.
Effectively, one can ask to what extent the quoted solutions/equations of motion can be distinguished experimentally by data from gamma ray bursts \cite{Bie01,Tak98}, ultra high energy cosmic rays \cite{Fin01} and cosmological microwave background fluctuations \cite{Mag05}.
In particular, in order to incorporate some discrete spatial and causal structures at the Planck energy scale, the action which leads to Lorentz invariance with an invariant energy scale ($E_{\P\l}$ or $m^{\2}/E_{\P\l}$) can be taken into account simultaneously with the action here proposed.
Ideally, the formalism we have discussed, in the sense analogous to that of the VSR, can be used to compare experiment and theory, as well as to extrapolate between predictions of different experimental measurements.
For fermionic particles with the dynamics described by the Eq.~(\ref{app10}), the phase space kinematic restriction is modified by a change in the relevant matrix element.
In this context, the subsequent application of our results concerns the verification of how the weak leptonic charged current $j_{\mu}$ must be modified to ensure its conservation so that we can examine the consequences for the experimental results were neutrino masses to have a purely, a dominantly or a perturbatively Lorentz-violating origin. 

For both of these analysis we have introduced the quantum deformations of the corresponding symmetries
such that the particular form of the spacetime coordinate transformations and the associated commutation relations remain covariant.
In general lines, the covariant formulation of the real space coordinate transformation as well as the determination of the commutation relations
provides an explanation of quantum mechanics based on scale relativity, which has not often been contemplated in modern physics, but can
be compared with previous results presented in the literature \cite{Sny47,Ban06}.

To conclude, there is a simple reason to be suspicious about the results here developed.
The supposed modifications at Planck scale do not respect the transformation laws of special relativity, according to which the energy and momentum transform in conformity with Lorentz transformations such that the Minkowski metric is preserved.
Since the Lorentz invariance is generally assumed to be a consequence of the relativity of inertial frames, one may then worry about whether the hypothesis here explored will be confirmed experimentally.
It could ruin the idea of isotropy of the Universe.
Moreover, it seems unlikely that Lorentz violation by itself, rather than neutrinos masses, can
explain observed neutrino oscillations.
However, we see this formalism as being part of a phenomenology of quantum gravity effects, as opposed to directly having
a fundamental significance, at the same time that, on the other hand, it implies three possible phenomenologically observable modifications to neutrino physics:
(i) modifications to the predictions for neutrinoless double beta decay \cite{Ver02}, (ii) some peculiar
modifications to the endpoint of the tritium beta decay \cite{Hol92} that is expected to be detected by
the next generation of endpoint experiments and (iii) small modifications to the oscillation picture due to
Lorentz-violating interactions that couple with the active neutrinos and eventually allow the complete explanation of neutrino data.

{\bf Acknowledgments}
The authors thank the professors C. O. Escobar and M. M. Guzzo for useful discussions and FAPESP (PD 04/13770-0 and PDJ 05/03071-0) for the financial support.

\end{document}